\documentclass[prl, reprint, superscriptaddress, amsmath,amssymb,aps]{revtex4-2}
\usepackage{graphicx}
\graphicspath{ {Figures/} }
\usepackage{dcolumn}
\usepackage[normalem]{ulem}

\usepackage{bm}
\usepackage[export]{adjustbox}
\usepackage{soul}
\usepackage{mathtools} 
\DeclarePairedDelimiter\bra{\langle}{\rvert}
\DeclarePairedDelimiter\ket{\lvert}{\rangle}
\DeclarePairedDelimiterX\braket[2]{\langle}{\rangle}{#1 \delimsize\vert #2}

\usepackage[utf8]{inputenc}
\usepackage[T1]{fontenc}
\usepackage{mathrsfs}
\usepackage{dsfont}

\usepackage{amssymb}
\usepackage{amsmath}
\usepackage{amsfonts}
\usepackage{amsthm}

\usepackage{hyperref}
\hypersetup{pdfpagemode=UseNone}

\newcounter{rem}
\def\tr{{\rm tr}}

\def\rho{{\varrho}}

\def\textbf#1{{\bf #1}}

\gdef\lvert{\delimiter"426A30C }
\gdef\rvert{\delimiter"526A30C }

\usepackage{xcolor}

\begin{document}

\title {Necessity of entanglement for the typicality argument in statistical mechanics}

\author{Pedro S. Correia}
\affiliation{Departamento de Ciências Exatas, Universidade Estadual de Santa Cruz, Ilhéus, Bahia 45662-900, Brazil}
\email{pscorreia@uesc.br}
\author{Gabriel Dias Carvalho}
\affiliation{Física de Materiais, Escola Politécnica de Pernambuco, Universidade de Pernambuco, 50720-001, Recife, PE, Brazil}
\author{Thiago R. de Oliveira}
\affiliation{Instituto de Física, Universidade Federal Fluminense, Av. Litoranea s/n, Gragoatá 24210-346, Niterói, RJ, Brazil}
\date{\today}

\begin{abstract}
Typicality arguments replace the postulated mixed‐state ensembles of statistical mechanics with pure states sampled uniformly at random, explaining why most microstates of large systems exhibit thermal behavior. This paradigm has been revived in quantum contexts, where entanglement is deemed essential, but no clear quantitative link between entanglement structure and typicality has been established. 
Here, we study pure quantum states with controlled multipartite entanglement and show that when entanglement grows with system size $N$, fluctuations in macroscopic observables decay exponentially with $N$, whereas if entanglement remains finite, one recovers only the classical $1/\sqrt{N}$ suppression. Our work thus provides a quantitative connection between entanglement structure and the emergence of typicality, demonstrating that entanglement is crucial for thermalization in small quantum systems but unnecessary to justify equilibrium in macroscopic ensembles. This unifies the classical and quantum foundations of statistical mechanics by clarifying precisely when and why entanglement becomes relevant.

\end{abstract}

\maketitle

\textit{Introduction} - Statistical mechanics aims to describe thermodynamic macroscopic behavior emerging from the microscopic dynamics of systems composed of a large number $N$ of particles. Tracking the positions and momenta of all constituents, that is, specifying a point $(q,p)$ in the $6N$-dimensional phase space, is not only impractical, but also fails to yield insight into macroscopic observables. Solving the full microscopic dynamics is likewise infeasible. To address this, statistical mechanics introduces probabilistic descriptions over microscopic configurations restricted by available macroscopic information such as total energy $E$.

The standard assumption for an isolated system with fixed energy is a uniform probability distribution over all accessible microstates -- the microcanonical ensemble. If the system is weakly coupled to a thermal bath, the probability of a configuration depends only on its energy, leading to the canonical ensemble. Remarkably, this ensemble-based approach -- the Gibbs prescription -- successfully recovers all equilibrium thermodynamic properties. It yields average values for macroscopic observables that agree with experimental measurements. Crucially, this identification is valid because ensemble fluctuations around these averages become negligible in the thermodynamic limit since they decay with $1/\sqrt{N}$.

A conceptual issue arises, however, since experimental observations concern the time evolution of a single system rather than a large ensemble. A common justification for using ensemble averages is provided by the ergodic hypothesis, which assumes that the system, over time, uniformly samples all accessible microstates. Although this idea is conceptually straightforward, establishing ergodicity in realistic systems is notoriously challenging, the timescales involved are too large, and the requirement is stronger than necessary - only a few macroscopic observables need to exhibit thermal behavior \cite{Lebowitz73,Singh13}.

The typicality argument provides an alternative explanation. As first proposed by Boltzmann and later formalized, this approach posits that for large $N$, most microscopic configurations yield macroscopic observables indistinguishable from the ensemble average. That is, for a typical configuration $(q,p)$, the observable $A$ satisfies $A(q,p) \approx \overline{A}$ for most $(q,p)$. This condition is far weaker than ergodicity and requires only that the system's dynamics not be restricted to a vanishing fraction of atypical configurations. In this view, thermal behavior emerges not from dynamics but from high-dimensional concentration phenomena, the law of large numbers. 

The typicality framework was rediscovered and rigorously developed in the quantum context in 2006, where it attracted significant attention due to the central role of entanglement \footnote{ The first, more general and mathematically rigorous work is \cite{Popescu06}. Almost at the same time, a less rigorous but more intuitive derivation was presented in \cite{Goldstein06}. Later, it was realized that many of the results were obtained before \cite{Lloyd2013, Gemmer09}, some even by von Neumann \cite{Goldstein10}.}. A pure state can yield a thermal mixed state for a subsystem in quantum systems, provided sufficient entanglement exists with the rest of the system. This mechanism allows the appearance of thermal probabilities without introducing them by hand, in contrast to the classical Gibbs approach. Entanglement plays the role of an intrinsic source of probabilistic behavior: tracing over part of a highly entangled pure state yields a mixed reduced state that resembles a thermal ensemble. However, the precise quantitative relationship between entanglement and typicality remains unclear. For example, it is unknown how the degree or structure of entanglement affects the likelihood of thermal behavior or the size of fluctuations. Furthermore, entanglement is not commonly observed in macroscopic systems, and typicality arguments have long existed in the classical, non-entangled setting.

In this work, we aim to clarify the role of entanglement in the typicality argument and its relevance for explaining thermal behavior. We consider random pure states with a controlled degree of multipartite entanglement and analyze the scaling of fluctuations in macroscopic observables. Our focus is on extensive observables -- sums of local quantities -- that characterize thermodynamic behavior. We show that, for fixed entanglement structure, fluctuations decay polynomially with system size, mirroring classical behavior. Only when the degree of multipartite entanglement increases with system size do fluctuations decay exponentially, enabling thermal behavior even in small systems. Thus, while entanglement is essential for understanding thermal features in small quantum systems -- such as those engineered in modern laboratories -- it is not required to justify ensemble averages in macroscopic systems as described by classical thermodynamics.

\textit{Textbook Approach} - In the standard formulation of Textbook Statistical Mechanics, one considers an isolated system with a fixed energy, or another macroscopic restriction, and assumes a uniform distribution over all compatible microstates, the equal a priori hypothesis. One can then consider a small part of the system (the subsystem) that exchanges energy with the rest (the bath) but weakly interacts with it. It is easy to show that under this approximation, the probability that the subsystem has an energy $E$ is given by $\exp{-(\beta E)}/Z$. From these probability distributions, one computes ensemble averages of macroscopic observables. These averages are only meaningful because fluctuations become negligible in the thermodynamic limit. For example, the relative energy fluctuation scales as $\frac{\sqrt{\overline{\Delta E^2}}}{\overline{E}} \sim  1/\sqrt{N} $, which vanishes for macroscopic systems. This scaling, a consequence of the law of large numbers, allows the identification of ensemble averages with thermodynamic quantities.

In the quantum case, for an isolated system with fixed energy, textbook treatments define the microcanonical ensemble as the maximally mixed state \(\Omega = \mathbb{I}_R/d_R\) on the constrained Hilbert space \(\mathcal{H}_R\). This identification relies not only on the postulate of equal a priori probabilities, often introduced through an entropy maximization over the populations \cite{Reichl}, but also on an additional random-phase assumption, according to which relative phases between different energy eigenstates within the same energy shell are uncorrelated and effectively average out \cite{Pathria, Huang}. As a consequence, although microscopically valid physical states may be coherent superpositions of energy eigenstates in \(\mathcal{H}_R\), the cancellation of random phases leads to a description in which only diagonal contributions survive. Under the usual assumption of weak coupling between a small subsystem $S$ and a large bath $B$, the reduced state $\Omega_S=\mathrm{Tr}_B[\Omega]$ takes the canonical form $\Omega_S=\exp(-\beta H)/Z$. In this ensemble-based description, coherence and entanglement present in individual pure states are effectively averaged out; decoherence or dephasing arguments are often invoked to motivate this loss of phase information.

As in the classical case, one may attempt to justify ensemble averages through ergodicity, now in a quantum setting. However, developing a quantum ergodic theory is even more challenging: notions like integrability and chaos are more subtle or ill-defined in quantum mechanics. Still, the scaling of fluctuations remains analogous, with the relative variance of macroscopic observables again scaling as $\sim 1/ \sqrt N$, thus allowing the identification of ensemble averages with thermodynamic quantities.

\textit{Typicality} - An alternative to the dynamical justification of ensembles is the concept of typicality. Rather than putting probability by hand, postulating a mixed state and relying on time evolution or ergodicity to justify it, typicality appeals to statistical properties of high-dimensional spaces: in systems with many degrees of freedom, most microstates compatible with macroscopic restrictions yield nearly identical values for macroscopic observables. That is, typical states appear thermal without invoking any dynamics.


This idea dates back to Boltzmann, but it was rigorously developed in the quantum setting in the mid-2000s. The central claim is that thermal behavior does not require assigning probabilities to microstates by hand. Instead of postulating that the physical state of the system is given by the mixed microcanonical ensemble $\Omega=\mathbb{I}_R/d_R$, one considers a single pure state $\ket{\psi} \in \mathcal{H}_R$, which is in principle fully specified, drawn from the set of states compatible with the macroscopic constraints. When the total system is divided into a small subsystem S and a large bath B, the reduced state is $\rho_S = \tr_B [\ket{\psi}\bra{\psi}]$. Typicality emerges when for most states $\ket{\psi} \in \mathcal{H}_R$ we have $\rho_S$ close to the canonical ensemble $\Omega_S = \tr_B [\Omega]$. To quantify "most states" we need to introduce a probability distribution or measure in $\mathcal{H}_R$ \footnote{The introduction of a probability distribution may seem to make the argument equivalent to the usual ensemble approach, where probabilities are postulated. This is in fact a subtle and rarely discussed point. However, the probability here is not introduced to represent physical ignorance, but to quantify the fraction of pure states with a given property; see Sec.~6.1.1 of \cite{Baldovin2025}.}. We will use the standard Haar measure, as in \cite{Popescu06,Goldstein06}, although more general distributions lead to similar results \cite{Reimann07}.

Mathematically, this closeness, or typicality, is quantified by the average trace norm distance \cite{Popescu06}:
\begin{equation}
\overline{||\rho_S - \Omega_S||_1} \leq \frac{1}{2} \sqrt{d_S \text{Tr}[\Omega_B^2]} \leq \frac{1}{2} \sqrt{\frac{d^2_S}{d_R}}
\label{canonty}
\end{equation}
The overline denotes averages over $\mathcal{H}_R$, $||.||_1$ the trace distance, and $d_S$, $d_B$, and $d_R$ denote the Hilbert space dimensions of the subsystem, the bath, and the restricted Hilbert space $\mathcal{H}_R$ of the total system, respectively. Here $\Omega_B = \mathrm{Tr}_S(\Omega)$ is the reduced microcanonical state of the bath, \footnote{Note that Typicality is valid even when the average reduced state $\Omega_S$ is not of the canonical form, as for strong coupling between system and bath}. Note that the average distance decays exponentially with the number of particles of the system $N$ since, in general, $\text{Tr}[\Omega_B^2] \sim 1/d_B$ or directly from $d_B \sim \exp(N)$. This implies that almost all pure states yield approximately thermal-reduced states when the bath is large \footnote{One can go further and use the Levi-Lemma to show that the probability for $\rho_S$ to be $\epsilon$ away from $\Omega_S$ decays exponentially with $\epsilon$ and $d_S \text{Tr}[\Omega_B^2]$.}. Also, in this framework, thermal probabilities are not postulated at the level of the physical state but emerge intrinsically from quantum entanglement. A pure global state yields mixed subsystem states due to quantum entanglement, making the statistical description an inherent feature of quantum mechanics, rather than an external assumption. Thus, this perspective has led to the view that entanglement is responsible, or even fundamental, for the typical behavior; separable states, for example, would then be regarded as atypical (not thermal) states.

From the typicality result for subsystem states, one can also show that the expectation values of subsystem observables exhibit typical behavior. But, when considering an observable, one can have typicality even for the whole isolated system. In fact, in 2007 \cite{Reimann07} it was shown that for a generic observable $A$
when sampling $\ket{\psi}$ with Haar measure one has
\begin{equation}
\overline{\Delta \langle A \rangle^2} \equiv
\overline{\big(\langle A \rangle_\psi - \overline{\langle A \rangle}\big)^2}
\;\leq\;
4\,\|A\|^2\,\mathrm{Tr}[\Omega^2],
\end{equation}
with $\langle A \rangle_\psi = \bra{\psi} A \ket{\psi}$,
the overline denotes averages over $\mathcal{H}_R$ and
\(\|A\|\) is the operator norm \footnote{The original bound involves the spectral diameter \(\Delta_A = a_{\max} - a_{\min}\). Since \(\Delta_A \le 2\|A\|\) for any Hermitian \(A\), this inequality immediately implies \(\overline{\Delta A^2} \le 4\|A\|^2\,\mathrm{Tr}[\rho^2]\).}. 
We quantify the degree of typicality using variance and refer to it as the size of the fluctuations, as it indicates the probability of getting an atypical value for $\langle A \rangle_\psi$ \footnote{Eq.~2 quantifies the fluctuations in $\langle A \rangle_\psi$. It does not guarantee that the intrinsic quantum fluctuation in a given state is small. For that one should also show that $\langle A^2 \rangle_\psi - \langle A \rangle_\psi^2$ is typically small or close to the ensemble values. This aspect of intrinsic quantum fluctuations is rarely discussed explicitly, but as shown in \cite{Lloyd2013}, it is possible to infer from Eq.~2 that they are of the same order as the fluctuations in $\langle A \rangle_\psi$.}. Again, usually the upper bound will decay exponentially with $N$, since $\text{Tr}[\Omega^2] $ usually decreases with $\exp{(N)}$. Thus, for macroscopic systems, we have Typicality for the microcanonical ensemble, for the whole system, at the level of observables.

These observations lead to two key aspects in the quantum scenario. First, Typicality can manifest at the state level, but only for subsystems or at the level of observables for the whole system; such differences have been discussed in \cite{Goldstein15, Goldstein17, Mori_2018}. Second, the magnitude of the fluctuations decays exponentially with $N$, much faster than the $\sim 1/\sqrt{N}$ obtained by the textbook approach and needed to explain the emergence of the thermal equilibrium behavior of macroscopic systems. Note also that for extensive observables $A_N$, $||A_N|| \sim \text{poly}(N)$, even the absolute fluctuations can become negligible and, much faster than needed to recover the textbook result. In the case of non-local observables, the relative fluctuation would still decay, something not expected from the textbook results.

However, the necessity of entanglement in typicality arguments is not fully understood. Although entanglement enables the emergence of mixed states from pure ones, there is no established quantitative relationship between its amount and the degree of typicality, besides a few related works \cite{Garnerone10, Correia24}. Neither is its role clear in the typicality of global observables. In fact, already in \cite{Reimann07} it is mentioned that ``our present findings suggest that the main root is the typicality property of the entire compound, which is in turn not entangled with any further system.'' Moreover, since macroscopic systems are rarely found in highly entangled states, it is natural to ask: How much entanglement is required to explain thermal behavior?

Although Typicality can be established for states or observables, our view is that, to study the emergence of thermal behavior in macroscopic systems, one needs only to consider observables, since these are what we measure in experiments and observe in daily life. Obtaining the full density matrix of a macroscopic system is infeasible -- it contains vastly more information than what is accessible in practice. Because experiments probe only a handful of macroscopic observables, we now turn to two concrete ingredients for our analysis: extensive observables and a complementary tool for dealing with entanglement, namely $K$-separable pure states.

\textit{Extensive Observables and $K$-Separable Pure States} - We focus on extensive observables, which are sums of local observables across the system. These quantities are physically relevant because they correspond to macroscopic measurements -- such as total magnetization, energy, or particle number -- and are accessible in experiments. Formally, we define an extensive observable $A_N$ on a system of $N$ subsystems as:
\begin{align}
    A_N = \sum_{l=1}^{N} \sigma^{(l)},
    \label{eq:AN}
\end{align}
where each $\sigma^{(l)}$ is a Hermitian operator that acts non-trivially only on the $l$-th subsystem. This structure ensures that $A_N$ captures additive contributions from local degrees of freedom. They are also experimentally realistic since their range of possible outcomes grows in direct proportion to $N$, reflecting the sum of local contributions from each subsystem \footnote{this point was made by Khinchin and is used by Landau, both were against the necessity of the ergodic hypothesis \cite{Singh13}. Assuming statistical independence of the system's parts, one finds that relative fluctuations of such extensive quantities scale as $1/\sqrt{N}$)}.

To isolate the role of entanglement in the emergence of thermal behavior, we introduce a restricted class of pure quantum states with controlled multipartite entanglement: the $K$-separable states. A pure state $\ket{\psi} \in \mathcal{H}_1 \otimes \mathcal{H}_2 \otimes ... \otimes \mathcal{H}_N$
is called $k$-separable if the system can be partitioned into $k$ disjoint blocks $B_1, B_2, ...., B_k$ such that 
\begin{equation}
  \ket{\Psi_K} \;=\;
  \bigotimes_{j=1}^K \ket{\psi_{B_j}},
  \label{eq:psiK}
\end{equation}
where $\ket{\psi_{B_j}} \in \mathcal{H}_{B_j} $ is a pure state on the degrees of freedom within block $B_j$. Crucially, there is no entanglement between blocks. For a clean scaling analysis, we set all blocks to have the same size $n_B = N/K$. This choice is compatible with homogeneous systems with local interactions, where no spatial region is distinguished and entanglement typically builds up in a roughly uniform manner \cite{hastings2006, eisert2010}.

This construction provides a tunable way to control the degree of multipartite entanglement. When $K=N$, each block is a single subsystem, so the global state must be fully separable (no entanglement). When $K=1$, the whole system forms a single block, so any global entanglement pattern is allowed, including fully multipartite entanglement. Intermediate 
$K$ values yield states whose multipartite entanglement is confined within blocks of size $n_B$.

We now consider the expectation value of the extensive observable $A_N$ in a $K$-separable state. Since each block $B_j$ contains a group of subsystems and the global state is a tensor product over blocks, the observable naturally decomposes as
$A_N=\sum_j^K A_{B_j}$ with $A_{B_j}=\sum_{l \in B_j} \sigma^{(l)}$. Then, the expectation value over the full state factorizes as:
\begin{equation}
\langle A_N \rangle_{\Psi_K} = \sum_j^K \langle A_{B_j} \rangle_{\psi_{B_j}}
\label{eq:expecAN}
\end{equation}
This decomposition will allow us to analyze how fluctuations scale with $N$ as a function of the entanglement structure -- our focus in the next section.

\textit{Typicality with Limited Entanglement} - We now examine how the typicality of extensive observables is affected when the entanglement in the global state is limited. To do so, we consider ensembles of random pure states that are $K$-separable, i.e., composed of $K$ entangled blocks of size $n_B=N/K$ and with no entanglement between them. Within each block $B_j$, the state $\ket{\psi_{B_j}}$ is independently sampled according to the Haar measure. This set of states has at most $n_B$-multipartite entanglement, allowing us to control the degree of multipartite entanglement and assess how it affects the decay of the fluctuations. 

We begin by analyzing the average expectation value. Using the expression for the expectation value in Eq. (\ref{eq:expecAN}) and considering the independence of the probability distributions in each block, we have:
\begin{align}
    \overline{\langle A_N \rangle}_{\Psi_K}
    &= \sum_{j=1}^K \overline{ \langle A_{B_j} \rangle}_{\psi_{B_j}} = \sum_{j=1}^K \tr(A_{B_j}\overline{\ket{\psi_{B_j}}\bra{\psi_{B_j}}}) \nonumber \\
    &= \sum_{j=1}^K \tr(A_{B_j} \rho_{B_j}) = \langle A_N \rangle_{\rho_K} \;,
\end{align}
where we defined $\rho_K$ as the average global state, expressed as the product state of the local average states $\rho_{B_j} = \overline{\ket{\psi_{B_j}}\bra{\psi_{B_j}}}$ in each block:
\begin{align}
    \rho_K = \bigotimes_{j=1}^K \rho_{B_j}.
\end{align}

Using the fact that $\ket{\psi_{B_i}}$ are independent, the variance of $A_N$ can be written
\begin{align}
    \overline{(\Delta{A_N})^2}
    &= \sum_{j=1}^K \overline{(\Delta{A_{B_j}})^2} = \sum_{j=1}^K \sum_{l \in B_j} \overline{(\Delta{\sigma^{(l)}})^2} \nonumber \\
    &= \sum_{j=1}^K \sum_{l \in B_j}
       \overline{
         \Bigl[
           \tr\bigl(\sigma^{(l)}\,(\psi_{l|B_j} - \rho_{l|B_j})\bigr)
         \Bigr]^2
       }.
\end{align}
Here, $\psi_{l|B_j}$ and $\rho_{l|B_j}$ are the reduced states of $\ket{\psi_{B_j}}$ and $\rho_{B_j}$, respectively, obtained by tracing out all subsystems in block $B_j$ except the $l$-th one, i.e., $\psi_{l|B_j} = \tr_{\{i \in B_j, i \neq l\}}[\ket{\psi_{B_j}}\bra{\psi_{B_j}}]$ and $\rho_{l|B_j} = \tr_{\{i \in B_j, i \neq l\}}[\rho_{B_j}]$.

Using the inequality $\left| \text{Tr}(A B) \right| \leq \| A \| \cdot \| B \|_1$, we can bound the variance as follows:
\begin{align}
    \overline{(\Delta{A_N})^2}
    &\leq \sum_{j=1}^K \sum_{l \in B_j}
       \|\sigma^{(l)}\|^2 \; \overline{\parallel\psi_{l|B_j} - \rho_{l|B_j}\parallel_1^2} \end{align}

To obtain our main result, we use a slight variation \footnote{It is direct to generalize the bound for $\overline{||.||_1}$ to $\overline{||.||^2_1}$ following the steps in Sec.VA of quant-ph/0511225} of the bound in Eq. (\ref{canonty}):
\begin{align}
    \overline{\parallel \psi_{l|B_j} - \rho_{l|B_j} \parallel_1^2} \leq \frac{1}{4} \dfrac{d^2}{d_B},
\end{align}
with \(d\) and $d_B$ the Hilbert space dimension of a single subsystem and the block, respectively. Using the above result, we can write
\begin{align}
    \overline{(\Delta{A_N})^2}
    &\leq \frac{N}{4} \|\sigma^{(l)}\|^2  \dfrac{d^2}{d_B} = \frac{N}{4} \|\sigma^{(l)}\|^2  \dfrac{d^2}{d^{n_B}}.
\end{align}
This is the main technical result of our work: a bound on the variance of extensive observables in $K$-separable ensembles, which depends explicitly on the block size $n_B$ and the total system size $N$. It provides a clear, quantitative link between entanglement structure and the suppression of fluctuations, allowing us to compare different regimes and analyze the emergence of thermal behavior. We turn to this analysis in the next section.

\textit{Discussion: Entanglement and Typicality} - The variance bound derived in the previous section reveals two distinct regimes corresponding to different entanglement structures. An exponential decrease in the fluctuations with the system size $N$ is only possible when the degree of multipartite entanglement - quantified by the block size $n_B = \tfrac{N}{K}$ - also increases with $N$. In other words, both $N$ and $n_B$ must grow together. If $n_B$ is fixed, fluctuations still decay but not exponentially. To better understand this, consider a system of qubits. 

\begin{figure*}[t]
\centering
\includegraphics[width=0.48\textwidth]{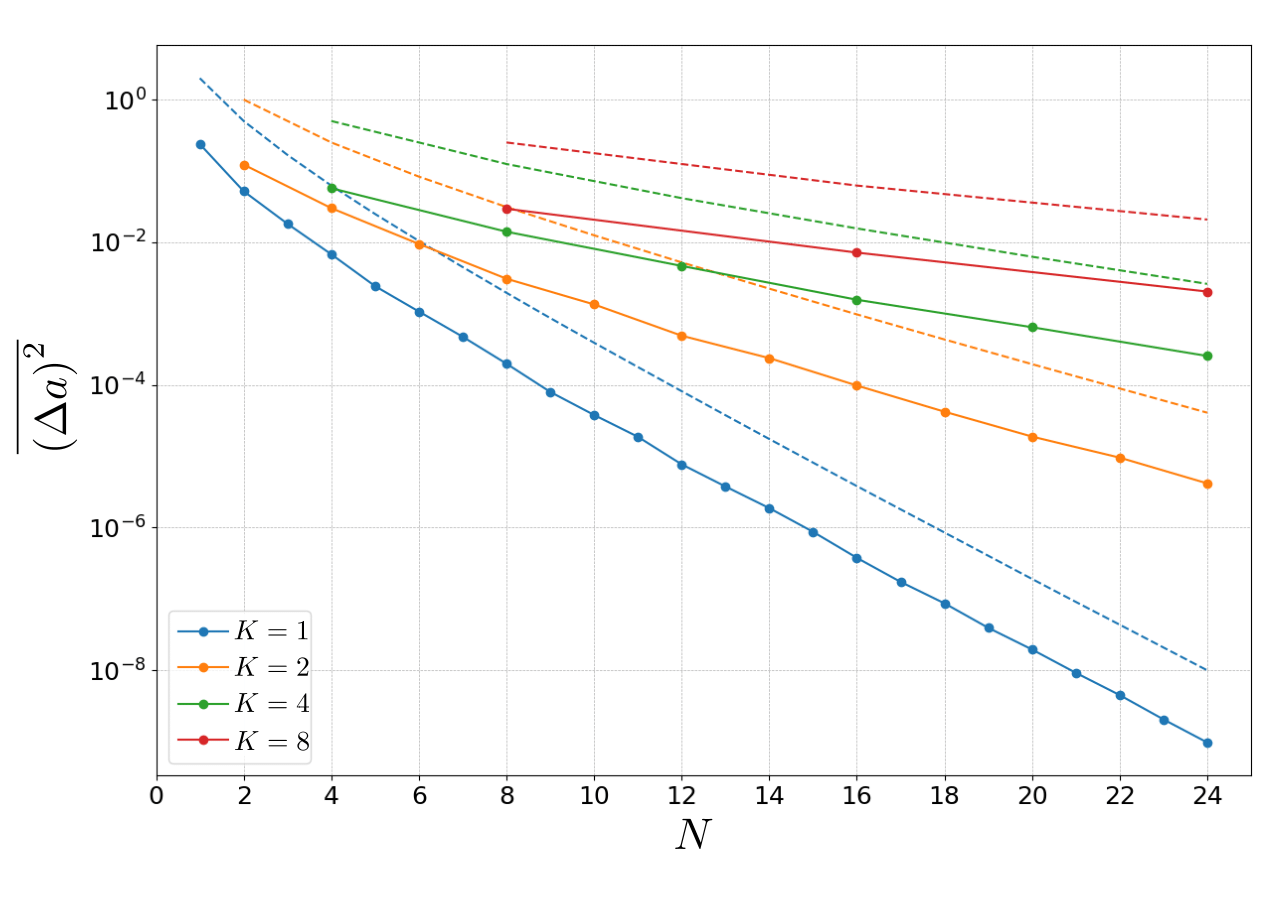}%
\hfill
\includegraphics[width=0.48\textwidth]{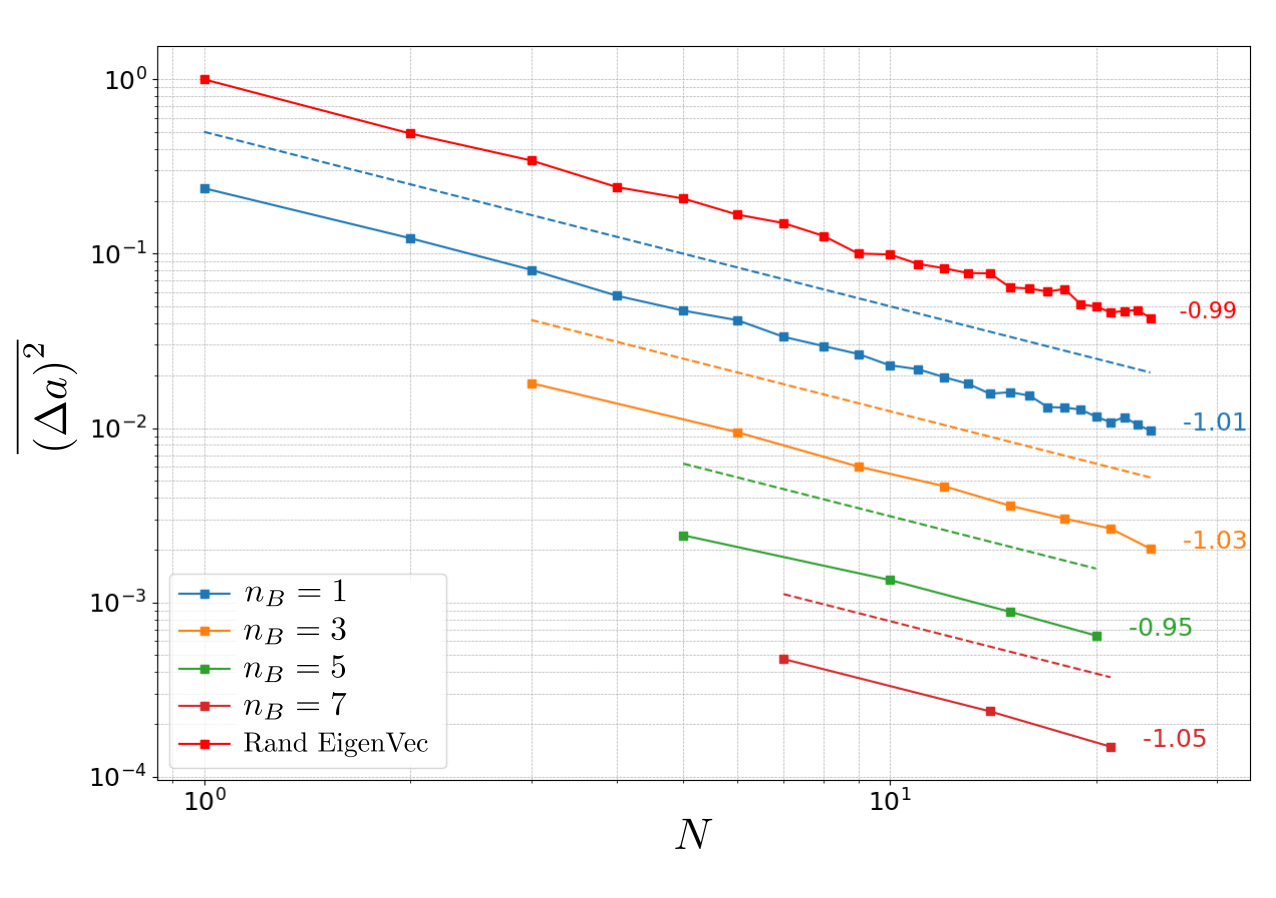}
\caption{ \textbf{Plots of $\overline{(\Delta a)^2}$ vs.\ system size $N$, illustrating the bound~\eqref{eq:bound-delta-a} under two partition scenarios.}
    Each marker averages 1000 Haar-random pure states generated with \textsc{QuTiP}~\cite{qutip}.
    \textit{\textbf{Left} (linear-$x$, log-$y$)}  --  the number of partitions $K$ is held fixed while the total number of qubits $N$ increases, so large-scale entanglement emerges. The variance decays almost exponentially and stays below the analytical upper bound (dashed lines), confirming the predicted scaling.
    \textit{\textbf{Right} (log-$x$, log-$y$)}  --  the block size $n_B$ is kept constant, so $K$ grows with $N$. In this case, the variance follows the expected $1/N$ behaviour, with fitted slopes close to $-1$. For standard quantum statistical mechanics reference, data for random eigenvectors (“Rand EigenVec”) of the observable show the same scaling.}
    
\label{fig:entanglement-typicality}
\end{figure*}

For qubits ($d = 2$), the blocks have dimension $d_B = 2^{N/K}$, and the operators $\sigma^{(l)}$ are Pauli operators with $\|\sigma^{(l)}\| = 1$. The variance bound then becomes:
\begin{align}
    \overline{(\Delta{A_N})^2}
    \leq \dfrac{N}{2^{N/K}}.
\end{align}
As noted earlier, macroscopic behavior's emergence depends on the relative fluctuations' decay. However, when the observable’s expectation value is zero, as often happens, relative fluctuations become ill-defined. To address this, we consider a density, intensive version of the observable, defined as $ a \coloneqq \frac{A_N}{N}$, and analyze its absolute variance:

\begin{align}
    \overline{(\Delta{a})^2} = \dfrac{\overline{(\Delta{A_N})^2}}{N^2}\leq \dfrac{1}{N}\dfrac{1}{\,2^{N/K}}.
    \label{eq:bound-delta-a}
\end{align}

As derived above, the exponential decay of the fluctuations in $a$ only occurs when $n_B$ grows with $N$, i.e., when the multipartite entanglement increases. If instead
$n_B$ is held fixed, the variance still decays (now as $1/N$) with a prefactor that depends on $n_B$, the degree of multipartite entanglement. Remarkably, even in the fully separable case ($N=K$), we recover the familiar $1/N$  scaling from the textbook approach.

This shows that entanglement is not required to justify using ensembles and explain the emergence of thermal behavior for macroscopic systems. Entanglement is needed only to explain the emergence of thermal behavior of small quantum systems, such as those studied in modern quantum experiments. In those systems, an exponential suppression of fluctuations is necessary to recover thermal behavior at small scales, and this is achieved via multipartite entanglement.

For a clean scaling analysis, Eq.~\eqref{eq:bound-delta-a} was written assuming equal block sizes $n_B = N/K$, which leads to the compact expression in Eq.~(13). If this assumption is relaxed, the corresponding bound involves a sum over block-dependent contributions, reflecting the different block sizes. Importantly, this modification does not affect the qualitative conclusions discussed above: the scaling behavior is still governed by how the typical (or smallest) block sizes grow with $N$, rather than by the specific details of the partition.

To illustrate our results, we perform numerical simulations for both regimes. We sample Haar-random pure states in the relevant Hilbert spaces using \textsc{QuTiP} \cite{qutip}. Each data point in Fig.~\ref{fig:entanglement-typicality} is averaged over 1000 samples. The left panel (linear-$x$, log-$y$ scale) shows where multipartite entanglement increases with $N$ (i.e., fixed $K$). The variance decays exponentially, in agreement with the theoretical prediction. Dashed lines indicate the upper bound, which captures the expected scaling. The right panel (log-$x$, log-$y$) shows the case of fixed multipartite entanglement (i.e., constant $n_B$). Here, the variance decays as $1/N$, with slopes near $-1$, consistent with the classical expectation. For comparison, we also include results for eigenstates of the observable, as in the textbook microcanonical approach based on uniform sampling of energy eigenstates, which follow the same scaling.

\textit{Conclusion} - The typicality framework, recently revived with an emphasis on entanglement, provides a compelling foundation for statistical mechanics and explains how irreversible thermal behavior emerges at macroscopic scales. Here, we have generalized that framework -- originally formulated for fully random pure states -- to ensembles of pure states with bounded multipartite entanglement. By deriving an explicit bound on the variance of extensive observables in $K$-separable states, we establish a quantitative link between entanglement and the suppression of fluctuations as system size grows.

Our results reveal two distinct regimes. If the block size $n_B$ (and hence multipartite entanglement) remains fixed, fluctuations decay only polynomially with $N$, reproducing the $1/N$ suppression of textbook statistical mechanics. Fluctuations decay exponentially only when the accessible multipartite entanglement scales with $N$, enabling typical thermal behavior even in small systems.

These findings resolve the longstanding question of when entanglement is essential for the Typicality argument for the foundation of statistical mechanics. Entanglement is not required to justify the emergence of thermal behavior in large macroscopic systems of our daily life, where fluctuations decay sufficiently even without entanglement. Entanglement becomes essential only in small, highly controlled quantum systems, such as the cold atoms and ions studied in many labs nowadays, and shown to present equilibration and thermalization. By unifying these classical and quantum perspectives, our work clarifies the scale‑dependent role of entanglement in the foundations of statistical mechanics.

As a future direction, it would be interesting to further quantify typicality in classical systems, as recently explored in \cite{Cattaneo25,Reimann2026}.

\textit{Acknowledgments - }This work is supported by the National Council for Scientific and Technological Development, CNPq Brazil (projects: Universal Grant No. 408990/2025-2, and 409611/2022-0). TRO acknowledges funding from the Air Force Office of Scientific Research under Grant No. FA9550-23-1-0092. 

\bibliographystyle{apsrev4-1}	
\bibliography{ref} 

\end{document}